\documentclass[10pt,conference]{IEEEtran}

\usepackage{balance}
\usepackage{url}
\usepackage[font=small]{caption}
\usepackage{amssymb} 

\usepackage{amsmath,amssymb,amsfonts}
\usepackage{algorithmic}
\usepackage{graphicx}
\usepackage{textcomp}
\usepackage{booktabs}
\usepackage{url}
\usepackage{xcolor}
\usepackage{diagbox}
\usepackage{xurl}
\usepackage{hyperref}
\usepackage{xspace}
\usepackage{subcaption}
\usepackage[htt]{hyphenat}

\usepackage[hang,flushmargin]{footmisc}
\def\BibTeX{{\rm B\kern-.05em{\sc i\kern-.025em b}\kern-.08em
    T\kern-.1667em\lower.7ex\hbox{E}\kern-.125emX}}

\AtBeginDocument{%
  \providecommand\BibTeX{{%
    Bib\TeX}}}
    
\newcommand{\ea}{{\em et al.}}
\newcommand{\eg}{{\em e.g.}}
\newcommand{\ie}{{\em i.e.}}

\usepackage{amsfonts}
\usepackage{algorithm}
\usepackage{textcomp}
\usepackage{booktabs}
\usepackage{makecell}
\usepackage[numbers,sort]{natbib}
\usepackage{diagbox}
\usepackage{endnotes}
\usepackage{enumerate}
\usepackage{graphicx}
\usepackage{hyperref}
\usepackage{listings}
\usepackage{longtable}
\usepackage{makecell}
\usepackage{multirow}
\usepackage{nicematrix}
\usepackage{tabularx}
\usepackage{tablefootnote}
\usepackage{tcolorbox}
\usepackage{xurl}
\usepackage{subcaption}

\tcbuselibrary{skins}
\newtcolorbox{mybox}[2][]{
top=0.15in,left=4pt,right=4pt,bottom=4pt,
fonttitle=\bfseries,
colbacktitle=gray,
colback=gray!5,
colframe=gray!40!black,
enhanced,
attach boxed title to top left={xshift=0em,yshift=-\tcboxedtitleheight/2},
boxed title style={size=small},
drop shadow={black!50!white},
title=#2,#1}

\begin{document}

\newcommand{\centralq}{%
	\vspace{0.5mm}
	\begin{center}
	\parbox[b]{0.72\columnwidth}{%
		\centering
		\emph{Are there performance or cost penalties associated with build tool downgrades?}
	}
	\end{center}
    \vspace{-0.5mm}
}

\title{{The Cost of Downgrading Build Systems} \\
\vspace{0.1em}
\LARGE A Case Study of Kubernetes}



\author{
\IEEEauthorblockN{Gareema Ranjan\IEEEauthorrefmark{1}, Mahmoud Alfadel\IEEEauthorrefmark{2}, Gengyi Sun\IEEEauthorrefmark{1}, Shane McIntosh\IEEEauthorrefmark{1}}
\IEEEauthorblockA{\IEEEauthorrefmark{1}Software REBELs, University of Waterloo, Canada; \IEEEauthorrefmark{2}University of Calgary, Canada\\
E-mail: \IEEEauthorrefmark{1}\texttt{\{granjan, gengyi.sun, shane.mcintosh\}@uwaterloo.ca}, \IEEEauthorrefmark{2}\texttt{mahmoud.alfadel@ucalgary.ca}
}}


\maketitle

\begin{abstract}
Since developers invoke the build system frequently, its performance can impact productivity. 
Modern artifact-based build tools accelerate builds, yet prior work shows that teams may abandon them for alternatives that are easier to maintain.
While prior work shows why downgrades are performed, the implications of downgrades remain largely unexplored.

In this paper, we describe a case study of the Kubernetes project, focusing on its downgrade from an artifact-based build tool (Bazel) to a language-specific solution (Go Build).
We reproduce and analyze the full and incremental builds of change sets during the downgrade period. 
On the one hand, we find that Bazel builds are faster than Go Build, completing full builds in 23.06--38.66 \% less time and incremental builds in up to 75.19 \% less time.
On the other hand, Bazel builds impose a larger memory footprint than Go Build of 81.42--351.07 \% and 118.71--218.22 \% for full and incremental builds, respectively.
Bazel builds also impose a greater CPU load at parallelism settings above eight for full builds and above one for incremental builds.
We estimate that downgrading from Bazel can increase CI resource costs by up to 76 \%.
We explore whether our observations generalize by replicating our Kubernetes study on four other projects that also downgraded from Bazel to older build tools.
We observe that while build time penalties decrease, Bazel consistently consumes more memory.
We conclude that abandoning artifact-based build tools, despite perceived maintainability benefits, tends to incur considerable performance costs for large projects.
Our observations may help stakeholders to balance trade-offs in build tool adoption.
\end{abstract}



\section{Introduction}

Build tools (\ie, tools that automate the process of transforming source code into deliverables) are a key development resource~\cite{hilton_2017}. 
Developers frequently invoke builds in their local environments as they modify the codebase.
Automated builds are also triggered by pull requests, build hooks, or nightly scheduled workflows. 
Since developers invoke builds frequently, build tool performance can boost or hinder developer productivity~\cite{hilton_2016, 8255774, 8530055, 10.1145/2568225.2568255,nejati2024understanding}. 

Artifact-based build tools, such as Bazel, Buck, Pants, and CloudBuild~\cite{10.1145/2889160.2889222}, have emerged to accelerate builds by enabling features like remote execution and caching. 
Unlike traditional tools (\eg, Make~\cite{feldman1979make}) or language-specific ones (\eg, Go Build), artifact-based tools maintain full control over the build process by explicitly tracking inputs, intermediate artifacts, and outputs. 
By constructing complete dependency graphs, these tools determine which tasks can be run in parallel and which artifacts can be fetched from a central cache, thereby minimizing repetitive build activity. 

Despite these benefits, it is not uncommon for projects to abandon artifact-based tools. 
For instance, in prior work~\cite{alfadel2024icse}, we found that 11.62 \% of projects that adopted Bazel (a popular artifact-based build tool) later downgraded to simpler build tools, such as Make or Go Build. 
These decisions are often driven by concerns over maintainability, contributor onboarding, and platform compatibility rather than build performance.
While prior work describes motivations for downgrades, abandoning artifact-based tools may incur performance penalties. 
In other words: are teams that downgrade sacrificing build speed and efficiency for maintainability?

In this paper, we investigate the performance and cost trade-offs of abandoning artifact-based build tools by asking:


\centralq

\noindent
To address this central question, we conduct a case study of Kubernetes---one of the most popular modern open-source systems for orchestrating containerized applications. 
Kubernetes is an ideal case for our study because the team downgraded to Go Build after over four years of using Bazel. 
During the transition, the project maintained both build tools concurrently, providing us with the unique opportunity to compare both tools in a real-world setting.


We replay full and incremental builds\footnote{Full builds run all build commands, whereas incremental builds only invoke commands that are (transitively) impacted by changed inputs.} for commits recorded before the downgrade event. 
Our experiments consumed approximately 3,402 computational hours---equivalent to over four continuous months of execution.
Through these experiments, we address the following Research Questions (RQs):

\begin{enumerate}[{\bfseries RQ$_1$}]
\item \textbf{How does the downgrade impact build duration?}\\
Build duration is a central factor in developer productivity. Bazel and similar tools are designed to reduce build time~\cite{zheng2024does}, but it remains unclear whether downgraded tools retain similar performance.
We observe that Bazel builds are consistently faster than Go Build, with a 23.06--38.66 \% speedup for full builds and up to 75 \% for incremental builds at higher parallelism levels.

\item \textbf{How does the downgrade impact resource usage?}\\
Performance gains may come at the cost of increased resource usage. 
We measure memory and CPU usage under various parallelism settings.
We observe that Bazel builds require significantly more memory—up to 351.07 \% more for full builds—and exert greater CPU load under higher parallelism. Go Build, in contrast, is more memory-efficient at lower parallelism levels.

\item \textbf{How does the downgrade impact the costs of continuous integration?}\\
Build-related costs in CI environments can be substantial, especially for large systems~\cite{hilton_2016, pinto2018work}.
We estimate these costs using resource pricing models from CI providers.
We observe that Bazel builds cost 22.62--39.14 \% less than Go Build for full builds and up to 75.92 \% less for incremental builds.
\end{enumerate}

While Kubernetes offers a unique opportunity to study build downgrades in a large-scale setting, its monorepo structure and scale may limit generalizability.
To mitigate this and assess the generalizability of our findings, we replicate the case study on four open-source projects drawn from the dataset of our prior work~\cite{alfadel2024icse}. 
Our replication demonstrates that although the performance gap between Bazel and Go Build narrows in smaller projects, Bazel consistently imposes a larger memory footprint than its replacement.

Our study implies that downgrading from artifact-based tools, such as Bazel, can introduce significant hidden costs, even in projects that made the change deliberately. While maintainability and onboarding concerns often motivate such downgrades, our results suggest that these decisions may come at a steep performance and financial cost, especially at scale.
For practitioners, our findings serve as a caution against assuming that simpler build tools will scale effectively without consequences.
Teams considering a downgrade should carefully evaluate performance and CI cost implications using project-specific benchmarks.
For researchers and tool builders, our study highlights a growing need for build systems that strike a better balance between performance and maintainability. Bridging this gap remains an open challenge, and an opportunity for the next generation of build tooling.

{\color{black}
To our knowledge, this is the first study to combine build duration, resource usage, and CI cost to quantify the penalties of downgrading from an artifact-based build system to a language-specific one. 
Prior work documented downgrade events~\cite{alfadel2024icse}, but did not measure their concrete costs.
Also, we do not intend to use the term `downgrade' to imply a migration to an inferior tool.
For homogeneous Go projects, Go Build can simplify onboarding and maintenance ~\cite{alfadel2024icse} while providing most of the key features of an artifact-based build tool.}

\vspace{4pt}
\noindent
\textbf{Data Availability.}
To foster replication, our data set, as well as our data collection and analysis scripts are available online.\footnote{\label{replication}\url{https://doi.org/10.5281/zenodo.15533609}}

\section{Study Design}
\label{sec:study_design}
This section provides our rationale for selecting Kubernetes as our main subject system (Section \ref{sec:study-design-subject-system}), and describes the Data and Environment Preparation (Section \ref{sec:dep}) and Replay Execution (Section \ref{sec:re}) stages in our study design.

\subsection{Primary Subject System}
\label{sec:study-design-subject-system}

We study the main project of the Kubernetes organization\footnote{https://github.com/kubernetes/kubernetes} because it offers a unique opportunity to investigate the cost and performance impact of shifting from an artifact-based build tool (Bazel) to a less feature-rich build one (Go Build).
The Kubernetes project initially used both Bazel and Go Build to define their build process in 2017.
Four years later, they downgraded to Go Build (on February 28, 2021\footnote{\url{https://github.com/kubernetes/kubernetes/pull/99561}}).
The remainder of this section describes the inclusion criteria that we considered and explains how the Kubernetes project satisfies each criterion.

\vspace{1pt}
\noindent \textbf{Criterion 1: Fair base of comparison.}
We set out to control for confounding factors to the largest extent possible when comparing the studied build tools.
To that end, the studied tools must build the same versions of the codebase in an identical computing environment.
The Kubernetes project enables us to perform a quantitative analysis of build tool performance, having officially maintained both Bazel and Go Build simultaneously for over four years,~\ie, we can use official project configurations to compare the performance of Bazel and Go Build on the same list of change sets (commits).

\vspace{1pt}
\noindent \textbf{Criterion 2: Usage and community adoption.} 
Studying the performance of build tools in projects with extensive community adoption ensures that our findings will have relevance to a real-world setting of significance.
Kubernetes, with 3.8K unique contributors, 115K stars, and 40.6K forks on GitHub, exhibits such widespread usage and popularity.
Moreover, the abandonment of Bazel by the main project of Kubernetes notably impacted other projects, leading to Bazel being abandoned by several subprojects within the same organization, as well as projects from other organizations~\cite{alfadel2024icse}. 

\vspace{1pt}
\noindent \textbf{Criterion 3: Detailed documentation.} To replicate historical builds exactly as performed by the project contributors, we rely on the official documentation available in the studied project repository. 
The detailed documentation of Kubernetes\footnote{\url{https://github.com/kubernetes/community/blob/fbef72efb89bb06b4df7e6858a88655736653042/contributors/devel/}} ensures that we use the same versions of tools and dependencies that have been recommended by the community, enabling reliable reproduction of builds in our experimental environment.

\subsection{Data and Environment Preparation (DEP)}
\label{sec:dep}
We set out to analyze the two build tools by collecting a list of commits where the performance of the tools can be compared (DEP1). In addition, we create a stable environment for building the collected list of commits using both tools (DEP2). 
Below, we describe each step.

\vspace{1pt}
\noindent \textbf{(DEP1) Select commits.}
We study the three-month period from December 1, 2020 to February 27, 2021, before the downgrade event on February 28, 2021.
During this period, 2,161 commits were recorded on the \texttt{main} branch of the Kubernetes project.
Since evaluating all 2,161 commits would require 3,781 computational days (or 10.35 computational years),\footnote{2,161 change sets × 12 settings × 3 repetitions × 2 build tools × 35 average compute minutes per change set} we sample commits on a daily basis by selecting the latest commit with a passing build status (according to the GitHub API\footnote{\url{https://docs.github.com/rest/commits/statuses}}) for each day in the studied period.
Applying this step selects 81 change sets for analysis.
Note that we exclude commits that only modified documentation or configuration files.
The median number of files changed among the studied commits is 22 (Q1=7, Q3=46).

\vspace{1pt}
\noindent \textbf{(DEP2) Prepare build environment.}
We prepare a build environment to replay builds by creating a Docker image that serves as an isolated environment for both Bazel and Go Build. 
The Kubernetes community maintains official documentation and hosts artifacts, such as the base Docker image\footnote{registry.k8s.io/build-image/kube-cross:v1.15.5-1} used for official Kubernetes builds.
We extend this base image with a layer that iterates over the list of studied commits by: (1) checking out the studied commit and (2) invoking the specified build type, while (3) collecting performance metrics. 
Since a central artifact cache has not been configured for Kubernetes, we do not configure our builds to use an artifact cache apart from the default settings of the two studied build tools.
We discuss the implications of this choice in Section~\ref{sec:validity}.


\subsection{Replay Execution (RE)}
\label{sec:re}

We evaluate full and incremental build scenarios because both play important roles in development~\cite{8094455}. Full builds are used to initialize developer environments and are a common failsafe solution when build dependencies are not carefully maintained \cite{miller1998recursive}. Moreover, many CI solutions invoke full builds for each job \cite{9311876}. Conversely, incremental builds quickly integrate and test only the artifacts that have changed with respect to either the previous build that was executed locally \cite{8094455}, or with respect to a shared cache of artifacts~\cite{10.1145/3497775.3503687}.



For build replay execution, we first configure the parallelism settings before running the builds (RE1). Next, we invoke full builds (RE2-F) and incremental builds (RE2-I) for the selected change sets.
While the builds are being performed, we collect performance metrics for analysis (RE3). 
We perform our analysis using two machines, each with an AMD EPYC 9174F 16-core processor, 128 GB RAM, and a PCIe storage device. 
Below, we describe each step in detail.


\vspace{1pt}
\noindent \textbf{(RE1) Configure parallelism.}
Build tools run independent tasks concurrently.
Thus, the number of available CPU cores can impact build performance. 
CI platforms also factor in the number of CPU cores allocated to a build job when computing service charges.
Since Kubernetes does not specify the level of parallelism used for running their builds, we perform builds with parallelism set to 1, 2, 4, 8, 16, and 32 to study how this setting impacts build performance. 
Throughout the paper, this setting is referred to as the \textit{parallelism setting}.

%


\vspace{1pt}
\noindent \textbf{(RE2-F) Perform full builds.}
In this step, we check out a studied commit and build it using a specified build tool (Bazel or Go Build) and parallelism setting, while recording performance metrics.
We then inspect the logs to determine the build outcome (success or failure).
Since we focus on commits known to have passed, any failing builds are assumed to be due to flakiness, and we re-invoke the build.

\vspace{1pt}
\noindent \textbf{(RE2-I) Perform incremental builds.}
To replicate incremental builds, we use the first commit in our list as a base and perform a full build.
Subsequent commits are built after applying their change sets to the prior copy of the codebase, thereby ensuring that only the updated artifacts are rebuilt.

To study whether full and incremental builds produce the same results, we include a hash-based output comparison.
We generate MD5 checksums of all files in the output directories of Bazel and Go.
For each examined commit, we compute the set of checksums once after the full build completes and again after the incremental build completes.

For Bazel, full and incremental outputs for the same commit were always byte-for-byte identical, with all MD5 checksums matching.
For Go, 93 commits (93.93 \%) also produced identical output files, with differences being observed in six commits.
These mismatches were confined to non-functional metadata, such as build IDs, debug paths, and timestamps in intermediate files.
When rebuilding these commits with \texttt{-trimpath} (to strip absolute paths) and \texttt{-buildvcs=false} (to omit VCS stamping), these mismatches were eliminated.
Thus, the mismatches that we observe in Go can be attributed to toolchain-specific metadata rather than correctness issues.

Finally, to ensure consistency, all builds were executed inside a Docker container of the same image, and the same commit was checked out for both full and incremental builds.



\vspace{1pt}
\noindent \textbf{(RE3) Prepare metrics.}
While each studied build is executing, we measure its duration (wall-clock time) using the Unix \texttt{time} command.
We also collect data on CPU usage (as a load percentage) and memory consumed by the build tools.
We collect these metrics using \texttt{docker stats} for each Docker container.
For each iteration of each build, we collect these metrics as time series data with a sampling frequency of four seconds.
Then, we analyze the descriptive median values to lift our analysis to the build iteration level. 
For incremental builds, we ensure that memory consumption is not affected by previous commits by subtracting the memory consumed at the end of the prior commit from that of the new one.

\section{Primary Study Results}
\label{sec:results}
This section presents the results of executing full and incremental builds of the Kubernetes project using Bazel and Go Build with respect to our two RQs. 
For each RQ, we first present our approach, and then discuss our findings.

\subsection*{RQ1. How does the downgrade impact build duration?}
\label{sec:rq1}

\subsubsection*{Approach}
We measure the wall-clock time required with each build tool. 
As described in Section~\ref{sec:study_design}, for each studied commit, we perform both full and incremental builds using Bazel and Go Build, with parallelism settings of 1, 2, 4, 8, 16, and 32. 
To counteract fluctuations in system load, we repeat each build three times and report the median value.
Since we observe an average standard deviation of 4.61 \% for full builds and 7.24 \% for incremental builds for the studied commits, we believe that our median build durations are stable enough to draw meaningful conclusions after three iterations.


\subsubsection*{Results} 
\begin{figure}
    \centering
    \includegraphics[width=\columnwidth]{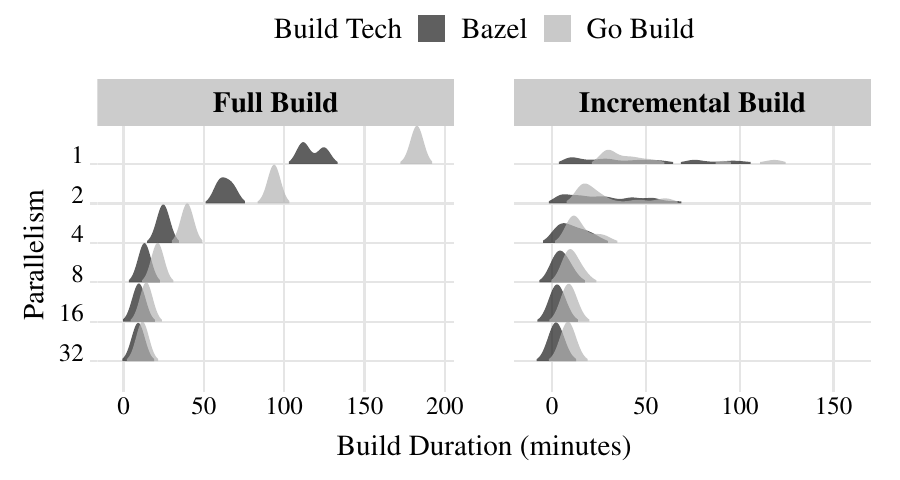}
    \caption{Median full and incremental build durations for Kubernetes.}
    \label{fig:full-time}
\end{figure}

Figure \ref{fig:full-time} shows the distributions of build durations.

\vspace{3pt}
\noindent
\textbf{\textit{Observation 1. Across all parallelism settings, full builds of Kubernetes executed using Bazel have significantly shorter durations than those using Go Build.}\label{observation1}}
In the baseline parallelism setting of one, Bazel has a median build duration of 112.40 minutes, whereas Go Build has a median of 182.70 minutes.
Although the gap between Bazel and Go Build narrows as the parallelism setting increases, Bazel maintains shorter build durations than Go Build.
Specifically, the median build durations for Bazel are 60.40, 24.19, 13.00, 9.46, and 9.13 minutes for parallelism settings of 2, 4, 8, 16, and 32, respectively.
In contrast, Go Build records median build durations of 93.76, 39.65, 21.23, 14.11, and 11.77 minutes.
Wilcoxon signed-rank tests (paired, two-tailed, $\alpha$ = 0.05\footnote{\label{bonferroni}We adjust $\alpha$ to $0.0083$ ($\frac{0.05}{6}$) using the Bonferroni method~\cite{heyde2001statisticians} to account for the six comparisons that we perform.}) indicate a statistically significant difference in build durations between the two build tools across all parallelism settings. 
Furthermore, the Cliff's delta effect size measure~\cite{gibbons2020nonparametric} indicates that the magnitude of the difference is large across all parallelism settings. 
The average overhead in build duration per commit when using Go Build compared to Bazel ranges between 23.06--38.66 \% across parallelism settings.



We analyze how each build tool processes individual targets across ten randomly selected commits to understand why Bazel achieves shorter overall build durations than Go Build.  
We examine build profiles and action graphs using the \texttt{-profile} and \texttt{-debug-actiongraph} options for Bazel and Go Build, respectively. These options reveal task execution times, dependencies, and sequencing, shedding light on the build strategy employed by each tool.
They also highlight where each tool spends its processing time.

We find that Go Build processes targets a median of 2.1$\times$--3.4$\times$ faster than Bazel for parallelism settings of 1 to 16, and a median of 6.1$\times$ faster at a parallelism setting of 32.
Although Go Build processes targets more quickly, Bazel achieves shorter overall build durations by decomposing tasks at a finer granularity.
Those finer-grained tasks can better leverage the resources that are available for the concurrent execution of independent tasks.


\vspace{3pt}
\noindent
\textbf{\textit{Observation 2. When parallelism is set to one or two, the difference in the incremental build durations between Bazel and Go Build is negligible; however, at parallelism settings greater than two, Bazel achieves significantly shorter build durations compared to Go Build.}\label{observation2}}
Bazel incremental builds have median durations of 48.01 and 27.19 minutes at parallelism settings of one and two, respectively, whereas Go Build has median durations of 44.63 and 22.27 minutes, respectively. 
Wilcoxon signed-rank tests (paired, two-tailed, $\alpha$ = 0.05\footref{bonferroni}) indicate that the difference is statistically significant at the parallelism setting of one, but insignificant at the parallelism setting of two. 
Moreover, Cliff's delta is negligible in both cases, suggesting that the magnitude of difference is minimal when parallelism is set to one or two.


As parallelism increases, the gap between incremental builds executed with Bazel and Go Build increases.
The differences become statistically significant for parallelism settings of four and above.
Cliff's delta for the parallelism setting of four is medium and is large for parallelism settings of eight and above.
The median build durations of Bazel at parallelism settings of 4, 8, 16, and 32 are 9.64, 4.27, 2.33, and 1.93 minutes, respectively.
In contrast, Go Build achieves median build durations of 12.71, 9.43, 8.47, and 8.13 minutes at the same parallelism settings.
Thus, the average overhead for abandoning Bazel is -3.69, -7.51, 29.17, 55.73, 69.84, and 75.19 \% at parallelism settings of 1, 2, 4, 8, 16, and 32, respectively.
Indeed, Bazel is particularly beneficial for incremental builds when parallelism is set above two. 

Analysis of the action graphs reveals that parallelization is also a key factor in speeding up incremental builds as well.
For individual targets, Go Build is 2.3x--3.2x faster than Bazel for the parallelism settings of 1 to 8.
At parallelism settings of 16 and 32, Go Build provides a speedup of 4.0x and 5.9x, respectively.
Thus, despite processing targets more slowly, parallelization helps Bazel achieve shorter overall build durations when parallelism is set to four or more.


To explore how caching mechanisms in Bazel contribute to its efficiency in incremental builds, we further analyze the time spent in each build phase using the \texttt{analyze-profile} option.
Bazel's build process is divided into \textit{launch}, \textit{init}, \textit{loading and analysis}, \textit{execution}, and \textit{finish} phases.
In the loading and analysis phase, Bazel constructs a dependency graph, identifying necessary dependencies, while the execution phase performs the build actions. 

For full builds, Bazel spends 22.4--48.9 \% of its execution time in the loading and analysis phase, with 61.2--78.7 \% spent on execution, varying based on the parallelism setting. 
This workload shifts substantially for incremental builds, with Bazel spending 1.0--3.4 \% of execution time in the loading and analysis phase and 97.5--98.8 \% on execution. 
This shows that Bazel conducts extensive dependency analysis and writes a substantial amount to caches during full builds with the goal of accelerating subsequent (incremental) builds.
In contrast, Go Build follows a simpler, phase-less design~\cite{whitney2019distributed}, compiling source code directly based on \texttt{import} statements, without leveraging caching for incremental optimization.

\begin{mybox}{Answer to RQ1}
 Full builds executed using Bazel are significantly faster than Go Build ones, across all parallelism settings.
 Choosing Go Build over Bazel introduces an overhead ranging from 23.06--38.66 \%. 
 For incremental builds, the difference in build durations between Bazel and Go Build is negligible when parallelism is set to one or two; however, as the parallelism setting increases, Bazel consistently maintains shorter build durations than Go Build.
 Indeed, the overhead for choosing Go Build over Bazel reaches as high as 75.19 \% at the parallelism setting of 32.
\end{mybox}

\subsection*{RQ2. How does the downgrade impact resource usage?}
\label{RQ:rq2}

\label{sec:rq2}

\subsubsection*{Approach}
To analyze the impact on resource consumption, we monitor both memory consumption and CPU usage.
We execute both full builds and incremental builds for each studied commit, repeating each build three times with different parallelism settings.
We collect time series data for resource consumption for each such build run
using \texttt{docker stats}.
We then compile the time series data into an aggregate time series representing the build metrics for each commit grouped according to the build tool.
From these two sets of time series', we analyze both memory consumption and CPU usage for each commit at each parallelism setting.

\subsubsection*{Results}
\begin{figure}[t!]
  \centering
  \includegraphics[width=\columnwidth]{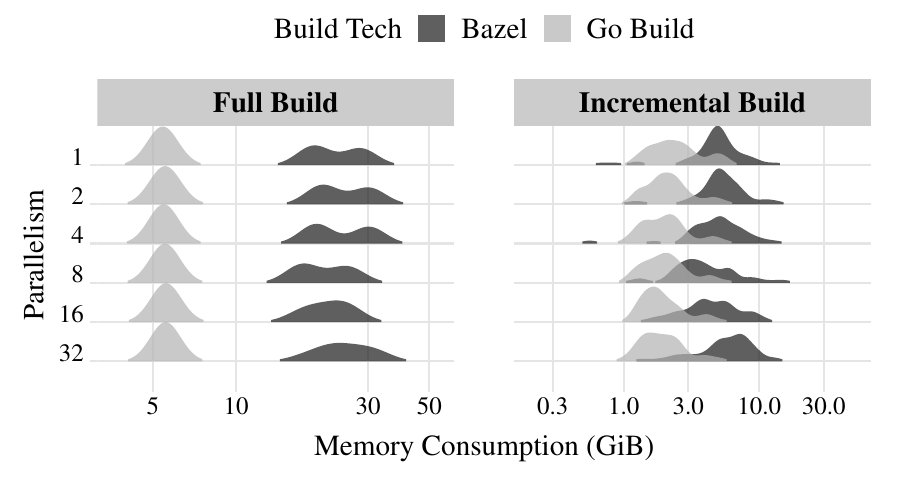}
  \caption{Median memory consumption (in GiB) for full builds and incremental builds for Kubernetes.}
  \label{fig:memory}
\end{figure}

\begin{figure}[t]
  \centering \includegraphics[width=\columnwidth]{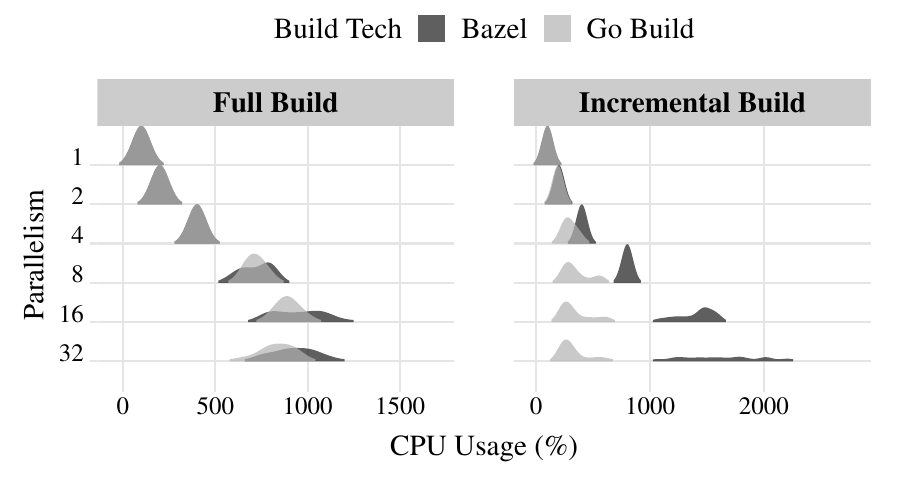}
  \caption{Distribution of median values of CPU usage (\%) for full builds and incremental builds for Kubernetes.}
  \label{fig:cpu}
\end{figure}

Figures~\ref{fig:memory} and~\ref{fig:cpu} provide an overview of the results.

\vspace{3pt}
\noindent
\textbf{\textit{Observation 3. Across all parallelism settings, Bazel has a significantly larger memory footprint than Go Build for both full and incremental builds.}\label{observation3}} 
Figure \ref{fig:memory} presents the distribution of the median memory consumption for full and incremental builds. 
For full builds, the median values range from 19.49--24.15 GiB for Bazel, and from 5.34--5.57 GiB for Go Build.
The median memory consumption of Bazel is 281.42--351.07 \% higher than that of Go Build across different parallelism settings.
Wilcoxon signed-rank tests (paired, two-tailed, $\alpha$ = 0.05\footref{bonferroni}) indicate that a statistically significant difference in memory consumption (with a large Cliff's delta) exists between Bazel and Go Build at all parallelism settings.

The higher memory consumption by Bazel can be attributed to two primary factors. 
First, its advanced caching mechanisms that increase memory consumption when Bazel optimizes build efficiency by caching build actions, including storing content hashes of source code.\footnote{\url{https://nicolovaligi.com/articles/faster-bazel-remote-caching-benchmark/}}
Second, Bazel's analysis of the build dependency graph~\cite{zheng2024does} results in a larger memory footprint for full builds. 
It uses available system memory to speed up dependency graph analysis, and cache build artifacts and dependencies as they are being created/downloaded. 

Observation 2 explains that Bazel has distinct phases of build execution. 
Bazel spends a considerable proportion of its execution time in the loading and analysis phase.
For full builds, Bazel spends an average of 22.4--48.9 \% in the loading and analysis phase, where it constructs and analyzes the build dependency graph and downloading (external) dependencies. 
This correlates with the larger memory footprint that is being generated by Bazel during full builds. 
In contrast, Go Build does not have a distinct and aggressive build graph analysis phase, and it resolves dependencies on-demand based on the \texttt{import} statements present in the code.

For incremental builds, the median memory consumption of Bazel ranges from 3.61--5.83 GiB, whereas Go Build uses between 1.8--2.39 GiB.
Bazel consumes 118.71--218.22 \% more memory on average than Go Build at any parallelism setting.
Wilcoxon signed-rank tests (paired, two-tailed, $\alpha$ = 0.05\footref{bonferroni}) indicate a statistically significant difference in memory consumption between the two tools with a large Cliff's delta at all parallelism settings. 

Similar to full builds, Bazel has a larger memory footprint than Go Build due to its caching and dependency analysis mechanisms. 
The memory consumption of both the build tools has reduced due to a decrease in the number of build actions and external dependencies being downloaded during incremental builds.
Bazel spends a smaller proportion of its execution time in its load and analysis phase, 1.0--3.4 \%, respectively.
Instead, Bazel spends more time on the execution of actual build tasks.



\vspace{3pt}
\noindent
\textbf{\textit{Observation 4. Go Build uses significantly more CPU resources than Bazel when running full builds at parallelism settings of four or less; however, the differences decrease when the parallelism setting is greater than four.}\label{observation5}}
Figure~\ref{fig:cpu} presents the distributions of CPU usage. 
For full builds, Go Build has higher CPU usage than Bazel at the parallelism settings of 1, 2, and 4.  
Wilcoxon signed-rank tests (paired, two-tailed, $\alpha$ = 0.05\footref{bonferroni}) indicate that Go Build incurs significantly more CPU usage at these parallelism settings, with a large effect size; however, this difference becomes insignificant with a negligible effect size when parallelism reaches eight. 
At the higher parallelism settings of 16 and 32 (\ie, with greater resource availability), CPU usage stabilizes for both tools, and Go Build has lower CPU usage than Bazel, with small and medium effect sizes, respectively.
Note that a parallelism setting of 32 is likely to be excessive for both tools, as the percentage of CPU usage decreases when the parallelism setting increases from 16 to 32.


We suspect that Go Build has greater CPU usage at low parallelism settings due to its adoption of \texttt{goroutines} for lightweight concurrent task execution within a single process, even on single-core systems~\cite{sottile2009introduction}. 
This model allows Go Build to handle multiple tasks with fewer processes~\cite{togashi2014concurrency}. 

To better understand the strategy of the two build tools when managing tasks, we analyze the number of processes used by each tool. 
Table \ref{table:pids} shows the median number of unique child worker Process IDs (PIDs) within Bazel and Go Build runs at each parallelism level.
The table shows that Go Build consistently creates fewer processes than Bazel. 
For full builds, as parallelism increases, Bazel creates more PIDs, while Go Build creates fewer. 
Go Build manages multiple tasks within a single process, which does not scale well (generating greater CPU usage) until the parallelism setting reaches eight.
At that point, the Go scheduler handles more tasks with fewer PIDs, reducing CPU usage as the parallelism setting increases. 
Conversely, for full builds performed using Bazel, the number of PIDs stabilizes from the parallelism settings of 2 to 16 and slightly increases at the parallelism setting of 32. 
With more CPU cores available beyond the parallelism setting of four, the parallel build tasks created by Bazel consume a larger amount of CPU resources.

\vspace{3pt}
\noindent
\textbf{\textit{Observation 5. Bazel incurs greater CPU usage than Go Build for incremental builds unless parallelism is set to one.
}\label{observation6}}
At the parallelism setting of one, Go Build uses more CPU resources than Bazel.
Wilcoxon signed-rank tests (paired, two-tailed, $\alpha$ = 0.05\footref{bonferroni}) indicate a statistically significant difference at this parallelism setting, with a medium Cliff's delta effect size.
For greater parallelism settings, Bazel uses more CPU resources than Go Build.
Statistical tests also reveal significant differences in the CPU usage at these settings, with a large Cliff's delta effect size.
Similar to full builds, CPU usage does not increase at the parallelism setting of 32 compared to that of 16 for both build tools.
Indeed, the median CPU usage for Bazel is 1451.3 \% at the parallelism setting 16 and 1641.93 \% at 32, whereas for Go Build, it is 300.64 \% and 290.22 \%, respectively.
This again suggests that the parallelism setting of 32 is not beneficial for incremental builds in this context.

\begin{table}[t]
  \centering
  \caption{The median number of PIDs across parallelism settings.}
  \label{table:pids}
  \vspace{0.3em}
  \begin{tabular}{@{}llrrrrrr@{}}
    \toprule
    \multicolumn{2}{c}{} & \multicolumn{6}{c}{\textbf{Parallelism Setting}} \\
    \cmidrule{3-8}
    \textbf{Build} & \textbf{Tool} & \textbf{1} & \textbf{2} & \textbf{4} & \textbf{8} & \textbf{16} & \textbf{32} \\
    \midrule
    \multirow{2}{*}{Full} & Go Build & 237 & 268 & 252 & 220 & 130 & 94 \\
                          & Bazel    & 579 & 687 & 691 & 689 & 686 & 718 \\
    \cmidrule{1-8}
    \multirow{2}{*}{Incremental}  & Go Build & 142 & 134 & 92  & 81  & 78  & 77 \\
                          & Bazel    & 617 & 673 & 684 & 696 & 691 & 711 \\
    \bottomrule
  \end{tabular}
\end{table}

We also study how the two build tools distribute and manage build tasks by studying the number of unique child worker processes they create (PIDs).
Table~\ref{table:pids} shows the number of PIDs created by each tool.
Bazel consistently creates more PIDs, indicating a larger number of worker processes. This typically leads to greater CPU usage for Bazel, especially during incremental builds, due to its fine-grained parallelism and the overhead associated with managing and scheduling numerous processes.
At the parallelism setting of one, the parallelism capabilities of Bazel are not being leveraged, resulting in less CPU usage than Go Build.
We attribute Bazel's higher CPU load at parallelism settings larger than two to its worker-based scheduling---a property of its architecture rather than its implementation language (Java).
Rewriting Bazel in Go would be unlikely to reduce its CPU usage.

\begin{mybox}{Answer to RQ2}
Bazel has a significantly larger memory footprint than Go Build, with overhead ranging from 281.42--351.07 \% for full builds and 118.71--218.22 \% for incremental builds.
In terms of CPU usage, Bazel effectively leverages available resources to accelerate its execution through concurrency, imposing a significantly higher CPU load for full builds when parallelism is eight or higher.
For incremental builds, Bazel imposes a higher CPU load unless its operation is serialized,~\ie, parallelism is set to one.
\end{mybox}

\subsection*{RQ3. How does the downgrade impact the costs of continuous integration?}
\label{sec:rq3}

\subsubsection*{Approach} 
%
The CI resource consumption cost of running builds is not directly proportional to the rate charged by CI providers.
Specifically, the cost of \emph{full builds} can be calculated as $N \times T \times r$, where:
\begin{LaTeXdescription}
    \item[$N$] represents the number of builds that are triggered each month. For Kubernetes, we derive the number of monthly full builds the project runs from Prow\footnote{\url{https://prow.k8s.io/}}---a CI platform used by the main Kubernetes project and its community to orchestrate build jobs. Since Prow only maintains the latest three months of build data, we could not retrieve the exact number of Kubernetes builds invoked during our selected study period (Section \ref{sec:dep}).
    Instead, we estimate $N$ using the median of 23 end-to-end full build executions per day (or 690 per month), that we find when we mine the most recent build data from Prow.\footnote{\url{https://prow.k8s.io/job-history/gs/kubernetes-jenkins/logs/ci-kubernetes-build}} 
    \item[$T$] represents the duration of each build. To estimate the duration of a full build, we calculate the median build duration for each parallelism setting based on the results of RQ1.
    \item[$r$] represents the rate of charge. Due to the absence of any pricing information for Prow, we approximate $r$ using the pricing model of GitHub Actions\footnote{\label{github-billing}\url{https://docs.github.com/en/billing/managing-billing-for-github-actions/about-billing-for-github-actions}}---a broadly adopted service provider for CI~\cite{9825792, kinsman2021software}. Table \ref{tab:pricing} shows that this pricing model factors in both the number of virtual CPUs allocated for build execution and the duration of the build process. Note that GitHub Actions does not provide a charge amount for processes running on a single CPU core. Hence, we exclude the parallelism setting of one from our cost analysis.
\end{LaTeXdescription}

\begin{table}[t]
  \centering
  \caption{GitHub Actions pricing based on virtual CPU count.}
  \label{tab:pricing}
  \vspace{0.3em}
  \begin{tabular}{@{}lrr@{}}
    \toprule
    \textbf{OS} & \textbf{vCPUs} & \textbf{Rate (USD/min)} \\
    \midrule
    Linux & 2  & \$0.008 \\
    Linux & 4  & \$0.016 \\
    Linux & 8  & \$0.032 \\
    Linux & 16 & \$0.064 \\
    Linux & 32 & \$0.128 \\
    \bottomrule
  \end{tabular}
\end{table}
To illustrate our approach using an example, consider the scenario of running full builds at the parallelism setting of two using Bazel. 
According to our findings in Section \ref{sec:rq1}, the median full build duration for this setting is $T = 60.40$ minutes. 
With the median number of monthly full builds derived from Prow data being 690, and the rate charged by GitHub Actions for this configuration being \$0.008 per minute, the monthly costs would be an estimated \$333.41 USD 
(\ie, $690 \times 60.40 \times \$0.008$).
Similarly, we compute the costs for other parallelism settings using their respective median build durations and the corresponding rates of charge in Table~\ref{tab:pricing}.

%
Although Kubernetes maintainers do not directly implement incremental builds for their CI environment, they still aim to reduce build times by using caching for Bazel, and skipping Go Builds if artifacts do not change.
Bazel supports the caching of intermediate and final build artifacts in a shared location.\footnote{\url{https://bazel.build/remote/caching}}
If that cache were stored in a persistent (external) location that CI builds could access, CI builds would essentially be performed incrementally.
Moreover, different CI services and tools such as GitHub Actions\footnote{\url{https://docs.github.com/en/actions/writing-workflows/choosing-what-your-workflow-does/}} and CircleCI\footnote{\url{https://circleci.com/docs/caching/}} can be used with tools to implement incremental builds by caching dependencies and artifacts. 
Recent research has also focused on the acceleration of builds by caching build environments and skipping unaffected build steps in CI, similar to an incremental build~\cite{9311876}.
Therefore, we perform a speculative cost analysis for the costs if a varying proportion of Kubernetes' CI builds had been incremental.

Incremental builds cannot run independently and depend on full builds as their starting point.
The need for full builds varies, affecting the frequency of subsequent incremental builds.
We estimate the cost by simulating various scenarios with different proportions of full builds supporting the incremental ones.
The monthly CI resource consumption costs for these scenarios can be calculated as $\left[ \beta \times T_{\text{full}} + (1 - \beta) \times T_{\text{inc}} \right] \times N \times r$, where $\beta$ is the proportion of times that a full build is run. 

After calculating the cost of both Bazel and Go Build, we calculate the relative difference in costs of builds using Bazel and Go Build with respect to Go Build.
A positive difference indicates that Bazel is more costly than Go Build, while a negative difference indicates the opposite.





\subsubsection*{Results} 

\begin{figure}[t]
    \centering
    \begin{subfigure}[t]{0.35\linewidth}
        \centering
        \includegraphics[height=4.4cm]{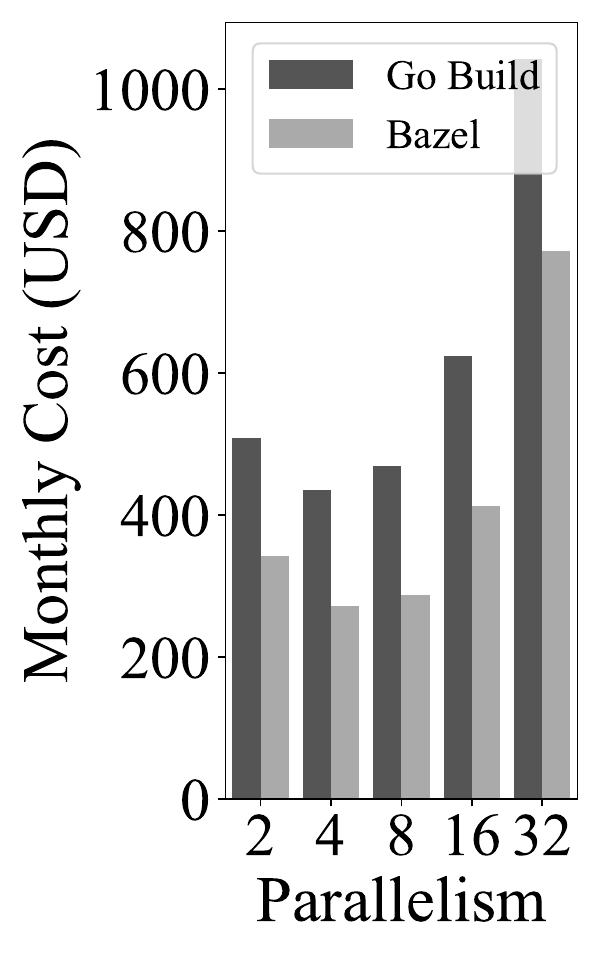}
        \caption{Monthly costs for full builds.}
        \label{fig:cost-full-build}
    \end{subfigure}
    \hfill
    \begin{subfigure}[t]{0.61\linewidth}
        \centering
        \includegraphics[height=4.4cm]{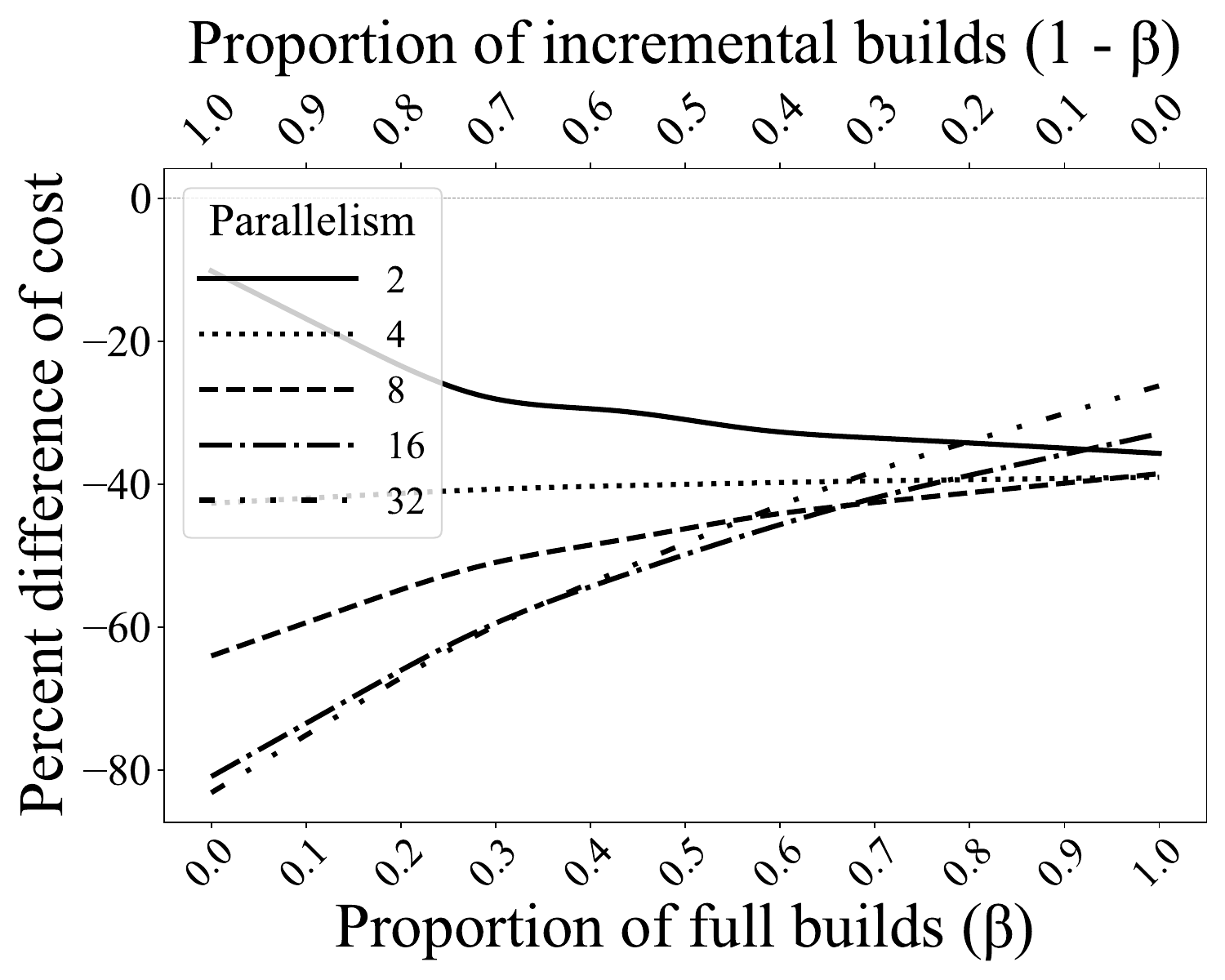}
        \caption{Cost differences for full vs. incremental builds.}
        \label{fig:cost-inc-build}
    \end{subfigure}
    \caption{Build cost analysis for Kubernetes using Bazel and Go Build.}
\label{fig:cost_inc_builds_2core}
\end{figure}

\vspace{2.5pt}
\noindent
\textbf{\textit{Observation 6. Bazel incurs lower costs than Go Build when executing both full and incremental builds for parallelism settings greater than two.}\label{observation7}}
Figure \ref{fig:cost-full-build} presents the estimated monthly costs for full builds of Kubernetes using Bazel and Go Build. 
From the figure, we observe that Bazel consistently incurs lower costs than Go Build, with CI cost savings ranging from 22.62--39.14 \% across parallelism settings.
Moreover, as the parallelism setting increases, so does the monthly cost for both build tools; however, an exception occurs when increasing the parallelism setting from two to four.
In that scenario, the CI cost decreases.
Despite incurring a higher per-minute price, the substantial reduction of build duration at the parallelism setting of four reduces the overall financial cost,~\ie, the time savings from the enhanced processing power outweighs the increased cost per minute.
From the results of RQ1 for full builds, we observe that when increasing the parallelism setting from two to four both Bazel and Go Build have the maximum speedup---2.5x for Bazel and 2.4x for Go Build. 



Figure~\ref{fig:cost-inc-build} illustrates the differences in cost (Y axis) between Bazel and Go Build as the proportion of full builds varies (X axis). 
From the figure, we observe that as the parallelism setting increases (plotted as different lines), Bazel becomes more cost-effective than Go Build.
For instance, at a parallelism setting of two and with the proportion of full builds $\beta=0.4$, Bazel is 21.95 \% more cost-effective than Go Build. 
This increases to 52.03 \% when the parallelism setting is 32.

The benefit of using Bazel is most pronounced at the highest observed parallelism setting of 32, when $\beta$ is as low as 0.01, resulting in savings of 75.96 \%.
In contrast, at the lowest studied parallelism setting of two, Bazel incurs more cost than Go Build when $\beta < 0.13$. 

Additionally, Figure~\ref{fig:cost-inc-build} shows that as the proportion of full builds increases, the cost difference between Bazel and Go Build narrows. This can likely be attributed to the diminishing optimization of the build process of Bazel when transitioning from incremental to more full builds, as discussed in Observations 1 and 2. 
Despite this convergence, negative percentages of cost difference indicate that Bazel incurs less cost overall.

Finally, we note that CI providers charge for CPU-minutes, so the cheapest configuration is not always the fastest.
For example, four CPU setting minimizes cost even though greater parallelism settings produce faster results.
This highlights that this analysis provides unique budget guidance that cannot be inferred directly from build duration measurements.

 

\begin{mybox}{Answer to RQ3}
Bazel is more cost-effective than Go Build across all parallelism settings, achieving savings from 22.62--39.14 \% for full builds and up to 75 \% for incremental builds. While its advantage decreases as the proportion of full builds increases, Bazel generally remains more cost-efficient.
\end{mybox}

\section{Analytic Generalizability}
\label{sec:gen-analysis}
Our study has thus far focused on the downgrade event within the main Kubernetes project, where Bazel was replaced by Go Build. 
To evaluate the generalizability of our findings, we expand the analysis along three complementary dimensions. 
First, we replicate our experiments on a larger and more recent time window of Kubernetes commits, to test whether our observations (O1--O6) remain valid (Section~\ref{repl:a}).
Second, we replicate our measurements on additional open-source projects to assess whether the observed trade-offs extend beyond Kubernetes (Section~\ref{repl:b}).
Finally, we conduct a qualitative comparison of build tools, based on documentation of Bazel, Buck, Pants, Go Build, and Maven, to understand which features are shared across artifact-based systems and which are specific to language-specific tools (Section~\ref{repl:c}).

{\color{black}
\subsection{Replication on Another Time Window}
\label{repl:a}
To assess whether our findings hold beyond the original three-month window (81 commits), we replicate our analysis on a larger and more recent period of Kubernetes development. We apply the same commit sampling strategy as we describe in Section~\ref{sec:study_design},~\ie, we select the last commit of each day. Starting from January 2023, we sequentially executed builds for each sampled commit using both Bazel and Go Build across the studied parallelism settings for both full and incremental builds.
For this analysis, we sample another 14 commits, which required approximately 14{,}000 minutes ($\approx$233 hours, more than nine days) of computation time. 

Overall, we find that this replication confirms our main trend,~\ie, Bazel achieves faster full and incremental builds across most parallelism levels. The only difference is that at the highest parallelism setting (32 cores), the performance gap between Bazel and Go Build becomes much smaller, with the two tools showing nearly comparable build times.
These runs also support our earlier observations for resource consumption,~\ie, the performance advantage of Bazel comes at the cost of a substantially larger memory footprint and heavier CPU load, whereas Go Build uses less memory, but shows different CPU utilization behavior without consistent performance gains.
In the appendix of our replication package,\footref{replication} we provide the detailed results, which are omitted here due to space constraints.

\vspace{1mm}
\noindent
\textbf{Comparison with original study.}  
Overall, our replication results largely confirm our original observations,~\ie, Bazel achieves shorter full and incremental build times than Go Build (O1, O2), but at the cost of higher memory and CPU usage (O3--O5).
The key difference lies in the maximum studied parallelism setting of 32 cores, where the performance gap between the two tools narrows substantially, with build times becoming nearly comparable.
In contrast, Section~\ref{sec:results} shows that Bazel maintains a clearer advantage even at high parallelism.
This shift may be attributed to the growth and restructuring of the Kubernetes codebase, which increases orchestration overhead in Bazel while allowing the simpler model of Go Build to close the gap.
Nonetheless, the overall conclusion holds; artifact-based systems like Bazel provide consistent performance benefits, while language-specific tools like Go Build remain more resource-efficient but less capable of sustaining speedups at scale.


}

\subsection{Replication on Other Projects}
\label{repl:b}

\begin{table}[t]
\centering
\caption{Results for smaller projects. H = Holds, NH = Not Hold}
\label{table:combined}
\resizebox{\columnwidth}{!}{%
\begin{tabular}{llcrcccc} 
\toprule
\textbf{Project} & \textbf{Lang} & \textbf{Alt Tool} & \textbf{Commits} & \textbf{Obs 1} & \textbf{Obs 2} & \textbf{Obs 3} & \textbf{Obs 4} \\
\midrule
\texttt{trillian}\tablefootnote{\url{https://github.com/google/trillian/pull/2743}} & Go & Go Build & 2,620 & NH & H & H & H \\
\texttt{emergent}\tablefootnote{\url{https://github.com/emer/emergent/commit/968d99}} & Go & Go Build & 724 & NH & NH & H & NH \\
\texttt{firedancer}\tablefootnote{\url{https://github.com/firedancer-io/firedancer/pull/157}} & C & Make & 576 & NH & H & H & NH \\
\texttt{entt}\tablefootnote{\url{https://github.com/skypjack/entt/commit/99f81e}} & C++ & Make & 1,193 & H & H & H & H \\
\bottomrule
\end{tabular}%
}
\end{table}

We replicate our measurements on four other open-source projects.
To select projects for our analysis, we revisit the dataset that we compiled in prior work~\cite{alfadel2024icse}.
We select projects with at least 500 commits before abandoning Bazel to ensure that each project has undergone a considerable amount of change so that the build tools can be compared.
Since we aim to characterize performance penalties, we exclude projects that configure the build tools to compile specific targets, rather than the entire codebase.

We find that nine candidate projects satisfy these criteria.
Of these, we could successfully replay the builds of the four projects that are listed in Table~\ref{table:combined}. 

Table \ref{table:combined} provides an overview of Observations 1--4 across the set of additional studied projects.
For example, Bazel's higher memory footprint (O3) holds consistently, while speedups (O1, O2) vary with project size.
Note that two of these projects perform a downgrade of the same type as Kubernetes (\ie, Bazel to Go Build), and two downgrade to a different technology (\ie, Bazel to Make). 

Unlike Kubernetes, changes to build specifications are needed to replay builds for this set of projects.
We modify the build specifications to address dependency issues due to the evolution of externally maintained packages. 
Prior work~\cite{tufano2017there, maes2022revisiting} also observed that replayed historical builds failed at a high rate of 71.36--91.26 \%, and a common problem was the evolution of external dependencies.
We provide these required changes as patches within our replication package.\footref{replication}

Since the selected projects are considerably smaller than Kubernetes, the build duration is shorter than one second for up to 22 \% of the studied incremental builds. 
As our evaluation setup is not sensitive enough to measure fluctuations in resource consumption for such short builds, we omit the analysis of incremental builds. 
Moreover, such short builds are unlikely to fluctuate enough to generate a meaningful difference in costs.
Hence, we do not perform our cost analyses (RQ3) on this set of projects. 

\begin{figure}
    \centering
    \includegraphics[width=\linewidth]{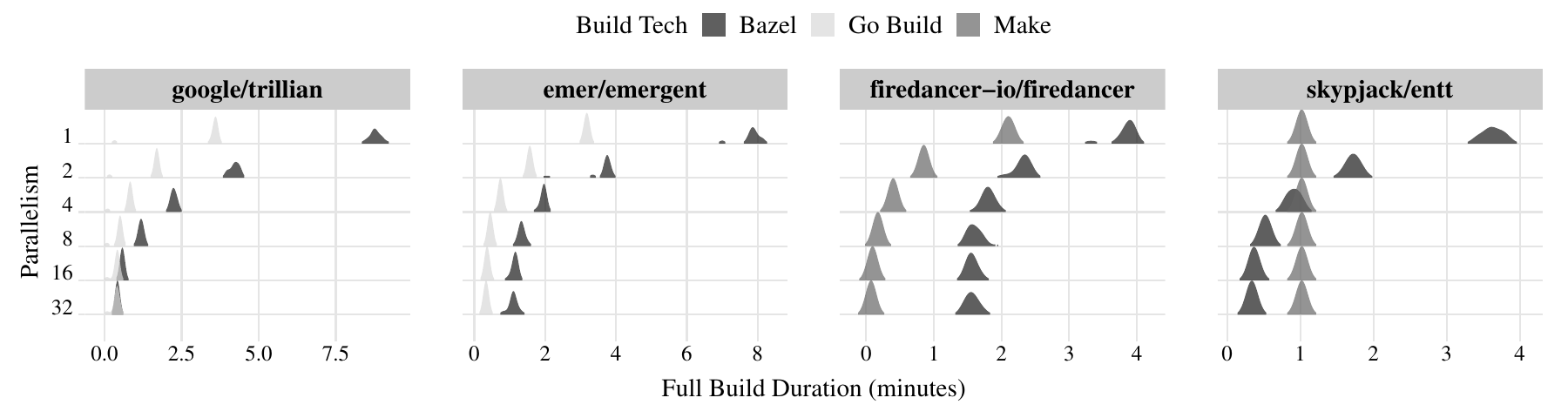}
    \includegraphics[width=\linewidth]{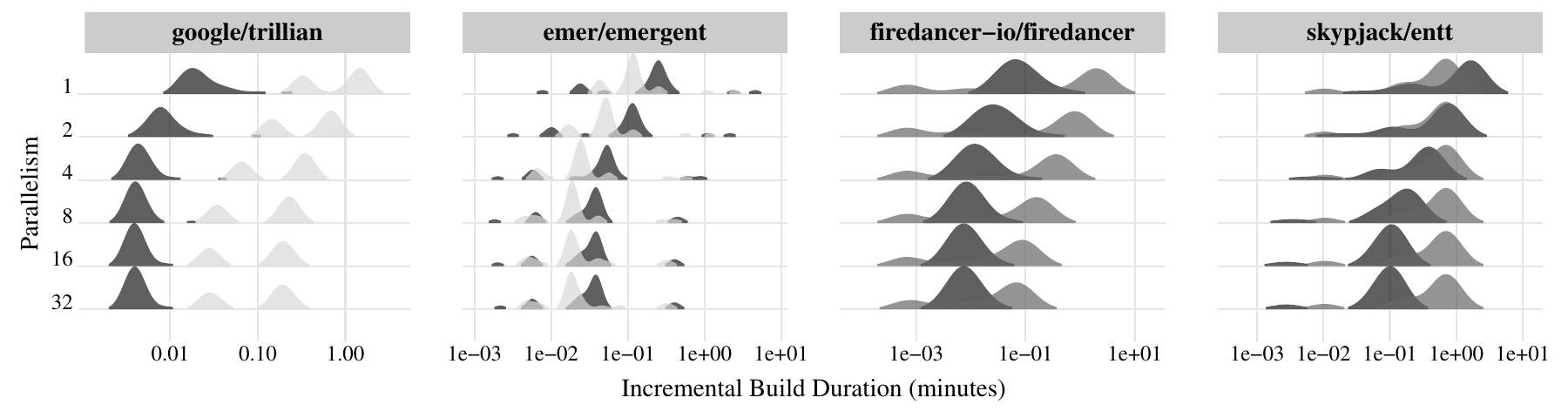}
    \includegraphics[width=\linewidth]{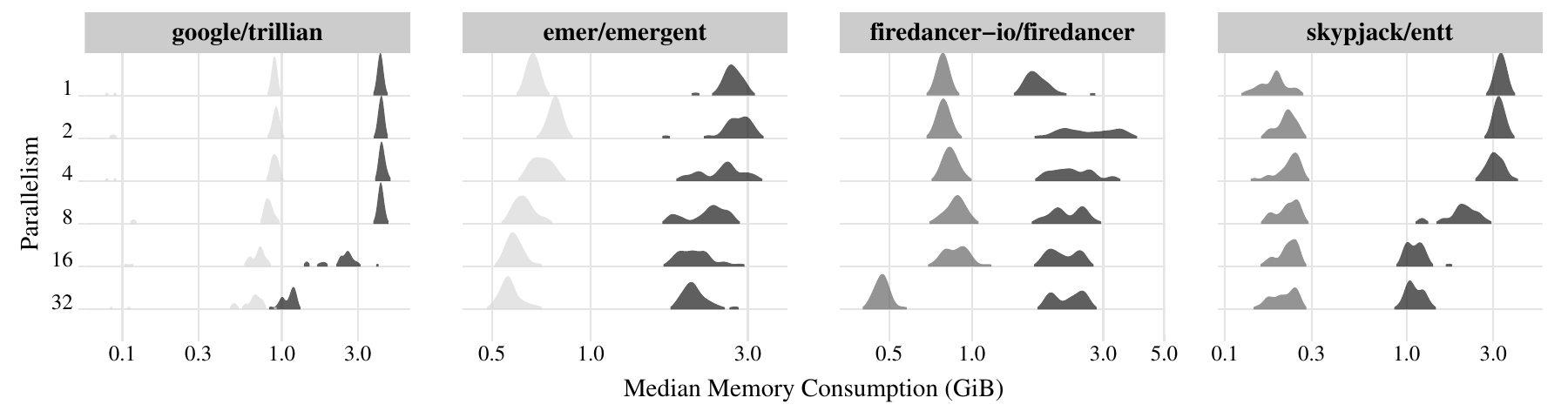}
    \includegraphics[width=\linewidth]{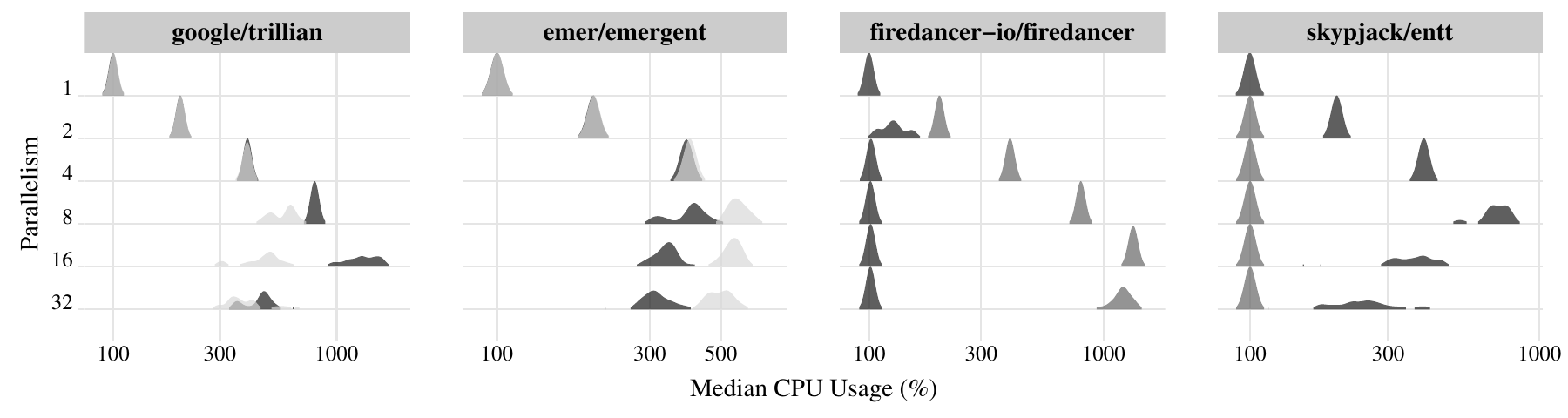}
    \caption{Performance metrics for analytic generalizability analysis.}
    \label{fig:gen-results}
\end{figure}

Below, we explore whether Observations 1--4 (Section~\ref{sec:results}) hold for these smaller projects.
Figure~\ref{fig:gen-results} provides an overview of the results for this new set of projects.

\vspace{3pt}
\noindent
\textbf{(Observation 1) Bazel has shorter full build durations across all parallelism settings.} This observation does not generally apply to smaller projects.
Wilcoxon signed-rank tests (paired, two-tailed, $\alpha$ = 0.05\footref{bonferroni}) indicate that builds executed using Go Build tend to have significantly shorter build durations than those executed using Bazel in both \texttt{google/trillian} and \texttt{emer/emergent}.
For projects downgrading to Make, the observation does not hold for \texttt{firedancer-io/firedancer}, but for \texttt{skypjack/entt}, we find similar trends in full build duration as Kubernetes at parallelism settings of four or more.

\vspace{2pt}
\noindent
\textbf{(Observation 2) Bazel has shorter incremental build durations except at low parallelism settings.} We note that this observation holds for three out of the four selected projects.
Incremental builds run using Bazel are significantly shorter at all parallelism settings in \texttt{google/trillian} and \texttt{firedancer-io/firedancer}. In \texttt{skypjack/entt}, similar to Kubernetes, the incremental build durations using Bazel are significantly shorter when the parallelism is set to four or above as indicated by Wilcoxon signed-rank tests,\footref{bonferroni} whereas for \texttt{emer/emergent}, the difference is insignificant between the two tools for incremental builds.

\vspace{2pt}
\noindent
\textbf{(Observation 3) Bazel has a larger memory footprint across all parallelism settings for full builds.} We note that this observation holds for all projects across all parallelism settings and Bazel consistently has a higher memory footprint. In fact, the Bazel community is developing tools to balance memory footprint with build flexibility as a trade-off.\footnote{\href{https://bazel.build/about/roadmap\#project-skyfocus}{https://bazel.build/about/roadmap\#project-skyfocus}}\textsuperscript{,}\footnote{\href{https://bazel.build/advanced/performance/memory\#trade-flexibility}{https://bazel.build/advanced/performance/memory\#trade-flexibility}}

\vspace{2pt}
\noindent
\textbf{(Observation 4) Bazel has higher CPU usage only at higher parallelism settings for full builds.} We observe different variations of this observation across projects.
In \texttt{google/trillian}, the trends of CPU usage are similar to Kubernetes until the parallelism setting of 16 with Bazel consuming more when parallelism is set to four or above. In \texttt{skypjack/entt}, we observe similar trends as Kubernetes; however, for \texttt{emer/emergent} and \texttt{firedancer-io/firedancer}, other tools tend to have a higher CPU usage than Bazel as the parallelism settings increase.

\vspace{3pt}
In summary, these replications show that while Bazel's memory overhead (O3) holds across all projects, speedups (O1, O2) vary with project size.
Thus, we believe that Kubernetes provides an example of a downgrade at scale, while smaller projects experience weaker penalties.
Moreover, the benefits of fine-grained task parallelization within Bazel are often outweighed by its overhead, particularly for full builds.
Go Build generally outperforms Bazel in terms of full build duration and memory usage.
Make, when used with its \texttt{-j} option, can also offer faster full builds and lower memory usage; however, Bazel retains an advantage in incremental builds across several projects.
CPU usage patterns vary across projects according to their build configuration, but Bazel consistently incurs higher memory consumption.

\subsection{Qualitative Comparison of Build Tools}
\label{repl:c}
To qualitatively compare build systems, we review their official documentation.
We conduct this analysis on Bazel,\footnote{\url{https://bazel.build/docs}} Buck,\footnote{\url{https://buck.build/setup/getting_started.html}} Pants,\footnote{\url{https://www.pantsbuild.org/dev/docs/introduction/welcome-to-pants}} Go Build,\footnote{\url{https://pkg.go.dev/cmd}}, and Maven\footnote{\url{https://maven.apache.org/guides/introduction/introduction-to-the-pom.html}}, focusing on how each tool structures builds, manages dependencies, caches artifacts, and exploits concurrency, to identify key features that drive build performance.
From this review, we identify five recurring architectural features that explain the trade-offs observed in our empirical results (O1–O6).

In our replication package,\footref{replication} we provide a table that summarizes these features, links them to our observations, and contrasts how they manifest in artifact-based with how they manifest in language-specific tools. 
The table is omitted here due to space constraints. 
Below, we describe each feature.

\vspace{1mm}
\noindent
\textbf{Directed action graphs and parallelism.} 
Artifact-based tools (Bazel, Buck, Pants) represent builds as Directed Acyclic Graphs (DAGs) of fine-grained tasks---a feature that enables extensive parallel scheduling.
This design contributes to why Bazel achieved shorter full builds (O1) and faster incremental builds (O2). Buck and Pants use similar DAG-based scheduling, whereas language-specific tools (Go Build, Maven) track dependencies more coarsely, and as such, cannot exploit parallelism as broadly.

\vspace{1mm}
\noindent
\textbf{Aggressive caching.} 
Bazel reuses build artifacts through both local and remote caches.
In our single-user experiments, only local caching is relevant, as Bazel avoids recomputing tasks when inputs remain unchanged, yielding significant incremental speedups (O2). Buck implements a similar mechanism using RuleKeys (hashes of inputs) to decide cache hits, while Pants supports both local and remote caches. By contrast, language-specific tools, such as Go Build and Maven, only support simple local caching (\eg, avoiding recompilation of unchanged packages or modules), which limits reuse and reduces incremental build gains.

\vspace{1mm}
\noindent
\textbf{Dependency graph overhead (memory).} 
Loading a fine-grained DAG will have a larger memory footprint than a coarse-grained DAG. Indeed, O3 shows that Bazel's footprint was up to 351 \% larger than Go Build's. Buck and Pants are likely to behave similarly because they must also load a fine-grained DAG, whereas language-specific tools resolve dependencies at a coarser granularity.

\vspace{1mm}
\noindent
\textbf{Worker process model (CPU).}
Artifact-based tools launch many worker processes to exploit parallelism, resulting in higher CPU utilization in both full (O4) and incremental (O5) builds. Buck and Pants share this model. By contrast, language-specific tools parallelize within a package or phase, using fewer processes overall. Thus, language-specific tools impose a smaller CPU load, but cannot scale concurrently to the same degree as artifact-based tools.


\section{Threats to Validity}
\label{sec:validity}

This section describes the threats to the validity of our study.

\noindent
\textbf{Construct validity.} 
A potential threat lies in how we define and measure performance impact.
We do not evaluate build hermeticity or CI/CD integration. These dimensions are important, but orthogonal to our goal of quantifying performance and cost penalties.
Our focus on duration, memory footprint, CPU usage, and CI cost follows prior work and reflects practitioner concerns about productivity and budgets.


\noindent
\textbf{Internal validity.}
Environmental factors like system load or network delays could affect our measurements.
To mitigate this, we use identical hardware, isolate builds in Docker containers, and ensure no competing processes run during execution.
We repeat builds three times; for Kubernetes, the standard deviation was 4.61 \% (full) and 7.24 \% (incremental), suggesting that our results are relatively consistent.

Another threat is bias in tool configuration.
We address this by selecting projects where both tools compile the full codebase and build identical targets.

We do not configure a remote cache for Bazel, as it mainly benefits repeated builds across users.
Since our focus is on sequential incremental builds by a single user, omitting the remote cache does not compromise our results.

\noindent
\textbf{External validity.}
Our primary subject is Kubernetes---a large-scale, resource-intensive system with a stable build pipeline and widespread adoption.
To improve generalizability, we replicate our analysis on four smaller projects that also downgraded from Bazel (Section~\ref{sec:gen-analysis}).

\section{Related Work}
\label{sec:related_work}
In this section, we position our work with respect to the literature on build migration (Section \ref{sec:rw-migration}) and build performance (Section \ref{sec:rw-performance}).

\subsection{Build Migration}
\label{sec:rw-migration}

Previous studies have explored migration between build tools~\cite{10.1145/3241625.2976011, mcintosh_build_maintenance, 6405267,yin2024developer}. 
For example,
McIntosh~\ea ~\cite{mcintosh_build_maintenance} studied projects that migrate upwards toward more feature-rich build tools (\ie, Ant to Maven and Make/Autotools to CMake), observing that they often pay off in terms of build maintenance activity (\eg, churn rate, logical coupling with source code).
Suvorov~\ea~\cite{6405267} mined the developer mailing lists of KDE and the Linux kernel to understand their build migration projects. 
They found that build migrations are often prolonged and error-prone, complicated by unclear requirements, communication bottlenecks, and the trade-off between performance gains and increased complexity.

Other studies have proposed methods to aid with build migrations~\cite{10.1145/2993274.2993279,gligoric2014automated}. For instance, Al-Kofahi~\ea~\cite{10.1145/2993274.2993279} introduced the AutoHaven platform that analyses and extracts the semantics of the GNU Autotools build tool, which can be further used to reduce manual migration effort.
Gligoric~\ea~\cite{gligoric2014automated} proposed a dynamic solution to automate the transition of build specifications to a different build tool.

Recent work has also demonstrated the tradeoffs in adopting the latest build technologies~\cite{alfadel2024icse,10.1145/3338906.3338922}.
For example, Alfadel and McIntosh~\cite{alfadel2024icse} studied
a phenomenon of projects migrating away from build tools like Bazel.
They found that developers using Bazel encountered technical, integration, and team coordination challenges, which have led them to downgrade to simpler, language-specific alternatives.
Similar to build tools, CI tools compete for adoption in the development marketplace. 

Our work takes inspiration from past work on build migrations and build tool downgrades, but differs by focusing specifically on the performance penalties that can realistically be associated with downgrading from a feature-rich build tool to a more traditional one. 

\subsection{Build Performance}
\label{sec:rw-performance}

Prior work has explored the impact of an efficient build process on developers~\cite{Rasmusson2004LongBT, hilton_2016,8255774,10.1145/2568225.2568255,8530055,hilton_2017}.
For example, Rasmusson \cite{Rasmusson2004LongBT} analyzed how prolonged build durations negatively impact development workflows and found that long build durations impact developer productivity and team spirit. 
Hilton~\ea \cite{hilton_2017} found that long build durations are a frequently referenced bottleneck, and developers argue that builds should take no more than 10 minutes to sustain a productive flow.
Maudoux and Mens~\cite {8255774} found that inefficient build systems in large projects result in longer build durations, directly impacting day-to-day developer productivity by increasing downtime between tasks.
A case study at Google~\cite{10.1145/2568225.2568255} also suggested that slow builds cause developers to lose context, reducing the number of changes they complete daily.

Given the substantial impact of slow builds on developer productivity, researchers have sought to understand and mitigate their causes~\cite{rogers,10.1007/s10664-019-09695-9}.
For example, Rogers~\cite{rogers} suggested strategies such as deciding on a maximum acceptable build duration, dividing build tasks into individual concurrent processes, and reducing test execution time to reduce build duration.
Ghaleb~\ea~\cite{10.1007/s10664-019-09695-9} examined factors affecting build durations. They found that beyond common factors like project size, team size, and test density, build duration can also be influenced by the configurations of artifact caching, the automated re-invocation of failing commands, and the timing of when builds are triggered.

%

Other studies have explored solutions to improve the performance of build tools by optimizing different aspects of the build process~\cite{
9311876,
8530055,
10.1145/3428212,
jendele2019efficient}.
For example, Gallaba~\ea~\cite{9311876} proposed a language-agnostic approach, Kotinos, to infer data from which build acceleration decisions can be taken. They
found that at least 87.9 \% of the 14,364 studied CI build records contained at least one Kotinos acceleration in their production setting.
Sotiropoulos~\ea \cite{10.1145/3428212} developed BuildFS to detect inconsistent build specifications for incremental builds, and helped achieve an average speedup of 74x when analyzing Make projects.
Lebeuf~\ea~\cite{8530055} designed BuildExplorer to debug and analyze build performance for distributed build tools with caching capabilities, such as CloudBuild.


Recent studies have explored various dimensions of the performance of artifact-based build tools, particularly Bazel
~\cite{jendele2019efficient, wang2021smart, zheng2024does}.
For example, Wang~\ea~\cite{wang2021smart} proposed a build target batching service (BTBS) that reduces errors in Bazel builds by optimizing memory usage and task execution, thereby improving the reliability of Bazel builds.
Zheng~\ea~\cite{zheng2024does} evaluated the performance of Bazel, focusing on its parallel and incremental build optimizations. 
Their study showed significant build speedups for long-build duration projects, with parallel builds achieving up to 12.8x improvement and incremental builds providing a 4.71x speedup. 
These findings highlight the potential of Bazel to improve build performance and highlight the need for better adoption and understanding of the latest build technologies.
Our study, however, analyzed the performance and cost implications of downgrading from a feature-rich build tool (\ie, Bazel) to a traditional one (\eg, Make and Go Build). 
We examine the consequences of abandoning Bazel in favour of a traditional build tool by comparing Bazel's performance under different parallelism settings with that of the other tool they adopted after abandoning Bazel.
Our performance measurement extends beyond build duration to include an examination of resource consumption, such as memory and CPU usage, as well as CI resource consumption cost for full and incremental builds.

\section{Final Remarks and Lessons Learned}
\label{sec:conclusion}

In this paper,  we study the impact of build tool downgrades through an empirical study to compare the performance (speed, computational footprint) of the prior artifact-based build technology (\ie, Bazel) with the less feature-rich replacement (\ie, Go Build). 
Furthermore, we explore the generalizability of our findings from Kubernetes to four smaller projects.
Distilled from our empirical observations, the following points summarise how our results translate into actionable advice for build‑system choice.

{\color{black}
\vspace{1pt}
\noindent
\textbf{Base decisions on project‑specific measurements.}
Before standardising on a tool, measure build duration, memory use and CI cost on representative commits; Bazel's cost advantage appears only at higher parallelism and with many incremental builds (O6). Hence, developers should benchmark their own projects under realistic workloads and compare the results to make an informed choice.

\vspace{1pt}
\noindent
\textbf{Opt for simplicity under resource constraints.}
When build agents have limited RAM or when builds run with low parallelism, the much lower memory footprint of Go Build and its simpler maintenance make it a sensible choice (O3). Hence, developers working with smaller codebases or constrained infrastructure are advised to use simpler, language-specific build tools like Go Build or Make.

\vspace{1pt}
\noindent
\textbf{Prioritize build speed when resources permit.}
For large projects where full‑build duration is the bottleneck or where many high‑parallelism incremental builds are executed, Bazel offers substantial speedups (O1 and O2). Hence, developers seeking shorter build times should select Bazel when sufficient memory and CPU resources are available.
A key trade-off, however, is between speed and CI cost. 
Our results (O6) show that four CPUs minimize CI cost while still providing substantial speedups, whereas higher parallelism further reduces build duration at greater financial expense. 
Developers should calibrate parallelism to their budget and workload priorities.
}

\balance
\bibliographystyle{IEEEtran}
\bibliography{ref}

\end{document}